\documentclass[preprint,12pt]{elsarticle}
\usepackage{amssymb}
\usepackage{multirow}
\usepackage{amsmath}
\usepackage{array}
\usepackage{nomencl}
\usepackage{eurosym}
\usepackage{xcolor}
\usepackage{subcaption}
\usepackage{booktabs}
\usepackage{url}
\makenomenclature
\journal{}

\begin{document}

\begin{frontmatter}

\title{A congestion-dependent imbalance pricing mechanism for regions allowing passive balancing}

\author{Hang Thanh Nguyen, Bart Van Der Holst, Phuong Hong Nguyen, Koen Kok} 

\affiliation{organization={Eindhoven University of Technology, Electrical Energy Systems Group, Department of Electrical Engineering}, 
            city={Eindhoven},
            postcode={5612 AP}, 
            country={The Netherlands}}

\begin{abstract}
Maintaining system balance becomes increasingly challenging as market design and grid capacity enhancement lag behind the growing share of renewables, requiring greater effort from both the transmission system operator (TSO) and the Balance Responsible Parties (BRPs). An actor can support balancing actively by bidding into reserve markets, or passively by adjusting its portfolio in line with system needs. In some countries, BRPs are incentivized to engage in passive balancing when their deviations support overall system stability. However, BRPs focus on profit maximization rather than minimizing portfolio discrepancies, which can cause simultaneous responses to price signals and create issues at the transmission–distribution interface. This research provides a two-stage stochastic model that captures BRP dynamic behavior and their impact on the grid under day-ahead and balancing market price uncertainty across three imbalance pricing mechanisms: the single, dual, and two-price. Then, a congestion-dependent imbalance pricing mechanism is proposed that maintains incentives for passive balancing while satisfying the grid constraint. A proof of concept is provided via the simulation with a part of the Dutch distribution grid. Results show that the proposed method mitigates the unexpected peak flow issue in congested areas while preserving passive balancing contributions from other BRPs in non-congested areas.
\end{abstract}

\begin{graphicalabstract}
\includegraphics[width=1\columnwidth]{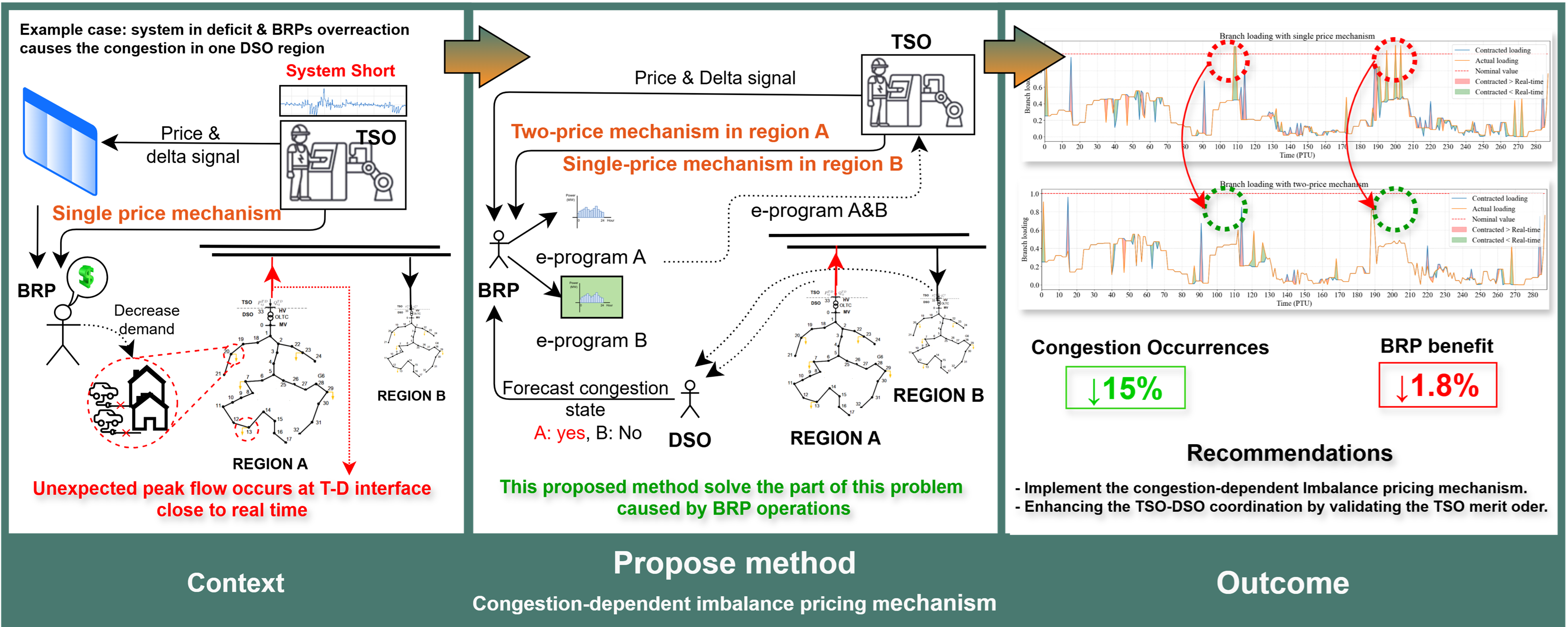} 
\end{graphicalabstract}

\begin{highlights}
\item A congestion-dependent imbalance pricing mechanism is proposed for regions that allows passive balancing by BRPs to address congestion in an economically efficient manner.
\item Providing a proof of concept via the simulation with a part of the Dutch distribution network.
\item Investigating the behavior of the balance responsible party (BRP) under different imbalance pricing mechanisms (IPMs) and considering its impact on the distribution grid.
\item Results show that the proposed mechanism effectively mitigates the peak flow issue in congested areas while still accepting the passive balancing in non-congested areas.

\item Providing recommendations for policymakers to enhance TSO-DSO coordination and consider the proposed imbalance pricing mechanism.

\end{highlights}


\end{frontmatter}

\nomenclature{TSO}{Transmission system operator}
\nomenclature{DSO}{Distribution system operator}
\nomenclature{BRP}{Balance responsible party}
\nomenclature{BSP}{Balancing service provider}
\nomenclature{PTU}{Program time unit}
\nomenclature{ISP}{Imbalance settlement period}
\nomenclature{IPM}{Imbalance pricing mechanism}
\nomenclature{RES}{Renewable energy source}
\nomenclature{DN/TN}{Distribution network and Transmision network}
\nomenclature{DA}{Day-ahead}
\nomenclature{EV}{Electric vehicle}
\nomenclature{VB}{Virtual battery}
\nomenclature{ASAP/ALAP}{As soon as possible and As late as possible}
\nomenclature{SOC}{Stage of charge}
\nomenclature{RT}{Real-time}
\nomenclature{SP}{Single price mechanism}
\nomenclature{TP}{Two-price mechanism}
\nomenclature{DP}{Dual price mechanism}
\nomenclature{V2G}{Vehicle to grid}

\nomenclature{$\pi$}{Probability of an scenario occuring}
\nomenclature{$\overline{E^{EV}_t}$}{The maximum value of the energy of the VB}
\nomenclature{$\underline{E^{EV}_t}$}{The minimum value of the energy of the VB}
\nomenclature{$E^{EV}_t$}{Energy of the VB at time t}
\nomenclature{$E^{arr}_{t}$}{Total energy from the EVs returning home at time t}
\nomenclature{$E^{dep}_t$}{Total energy from the EVs leaving home at time t}
\nomenclature{$E^{ALAP}_{n,t}$}{The minimum energy boundary is formed by the ALAP charging path}
\nomenclature{$E^{ASAP}_{n,t}$}{The maximum energy boundary is formed by the ASAP charging path}
\nomenclature{$N^{arr}_t$}{Number of EVs arrive at time t}
\nomenclature{$N^{dep}_t$}{Number of EVs depart at time t}
\nomenclature{$P^{\max}_k$}{The maximum charging rate of EV number $k$}
\nomenclature{$p_{charge_t}, p_{dis_t}$}{The maximum charging/discharging power at time $t$}
\nomenclature{$N_{parked}$}{The number of EVs parked at home}
\nomenclature{$E^{da}_t$}{The volume of energy bought from the DA market at time t on the next day}
\nomenclature{$\eta$}{The charging efficiency of the EVs}
\nomenclature{$\Delta t$}{Duration of an imbalance settlement period in an hour}
\nomenclature{$\lambda^{\uparrow}_{s_{rt},t}$}{The forecasted imbalance price for upward regulation at time step t, in scenario $s_{rt}$}
\nomenclature{$\lambda^{\downarrow}_{s_{rt},t}$}{The forecasted imbalance price for downward regulation at time step t, in scenario $s_{rt}$}
\nomenclature{$u$}{The binary variable helps prevent the surplus and shortage from happening at the same time}
\nomenclature{$\Delta E^{\uparrow}_t$}{Energy increases in real time compared to the energy bought from the day-ahead market}
\nomenclature{$\Delta E^{\downarrow}_t$}{Energy decreases in real time compared to the energy bought from the day-ahead market}
\nomenclature{$g,h$}{The binary variables used for ensuring the upward/downward imbalance price is applied to the right regulation state}
\printnomenclature

\section{INTRODUCTION}
\label{sec1}
Energy trading and wholesale markets are not new concepts, however, they remain relatively young compared to the electricity system and continue to evolve rapidly. Since 1996, five legislation packages have been introduced in Europe to promote the liberalization of the electricity sector, support the development of a new market design, and align with the EU's net-zero climate ambitions~\cite{18,third_package}. Liberalizing the electricity market brings economic benefits by lowering entry barriers for new participants and increasing competition, which is particularly emphasized in the Third Energy Package~\cite{third_package}. However, this increased competition also introduces greater complexity in maintaining real-time system balance. Besides, with the growing penetration of renewable energy sources (RES) and the ongoing electrification, system balancing becomes increasingly challenging due to the inherent unpredictability of RES and the volatility of electricity prices. As a result, system operators expect greater contributions from balance responsible parties (BRPs).
The BRP is financially responsible for the deviation from its own portfolio~\cite{30}, which is submitted to the transmission system operator (TSO) in the form of an e-program. To mitigate the deviation, BRP can (1) improve the forecast of power output accuracy beforehand, (2) buy/sell in the intraday market or via bilateral contract with other supplier/consumer to internally correct their deviation in real-time, or (3) leverage flexibility through coordination with other actors like virtual power plants, flexibility service providers~\cite{16, usef}. Meanwhile, the TSO performs the corrective actions and settles the BRP's deviation afterward using the imbalance price~\cite{7}. 
The imbalance pricing mechanism (IPM), which is the backbone of the balancing market, will thereby influence the behavior of the BRP. This pricing mechanism is the method implemented in the settlement period to determine the price applied for the actual deviation of the BRP from its prognosis. 
There is still no consensus on the point of view of the system operator among different countries and within a country itself at different times on choosing a suitable IPM for their system~\cite{13, 17, 18}. 
Some have concluded that allowing BRPs to optimize their passive balancing operations could contribute to their profits and effectively improve system stability, rather than inhibiting their influence~\cite {3,11}. BRPs can also create additional income streams by performing passive balancing~\cite{6,12,18}. 

However, alongside the economic benefits, the passive balancing action from multiple BRPs reacting simultaneously may lead to local congestion~\cite{thesis_Nobel}. This arises because BRPs are unaware of their own deviation and the actions of other BRPs and the TSO, while the TSO also lacks visibility into the current and future reactions of BRPs~\cite{8}. As a result, these uncoordinated actions counteract the re-dispatch for congestion management and cause an unexpected peak flow in the connected distribution grid if assets in the distribution grid are used for balancing purposes.

Recently, the congestion has become more and more severe in the distribution network in Nordic countries, the United Kingdom, Germany, and the Netherlands, etc.~\cite{conges_1,conges_2}. 
Particularly, the system operators in the Netherlands have observed various unexpected peak flows in the connection point between the transmission network (TN) and the distribution network (DN).
According to records from the TSO and the DSO, in the second half of 2023 to early 2024, many unexpected peak flows occurred at several substations lasting from 1 program time unit (PTU) to several hours, with the overload level reaching over 100\% of the nominal value.

There are several main causes of this situation. Firstly, this problem may come from the utilization of balancing services by TSO, provided by resources connected to the DN, without the DSO having visibility into those actions. Secondly, this issue might come from the inaccurate forecast of BRP or their internal balancing actions. Thirdly, multiple BRPs try to help the system via passive balancing actions.
The first cause can be addressed by enhancing coordination and transparency between the TSO and DSO, as demonstrated in our previous work~\cite{26}.
This study, however, focuses on the second and third causes: Congestion resulting from the BRP actions.
This situation will become increasingly serious both in frequency and severity as more distributed flexibility services are exploited and the pace of grid upgrades and policy changes does not keep up with the growth of electricity demand and electrification.

The above discussion poses a question: \textbf{How can passive balancing contributions from BRPs be effectively maintained and promoted without compromising the distribution grid constraint?} 
To explore this, our study focuses on the existing balancing market design, investigates the behavior of BRPs under different IPMs, as well as their impacts on the distribution grid. Thereby, a solution is proposed to better align BRP incentives with grid constraints.


\section{RESEARCH BACKGROUND}
\label{sec2}
\subsection{Electricity market in the Netherlands}
\subsubsection{Spot market}
Similar to most other countries with a liberalized electricity market, the Netherlands has both long-term and short-term electricity markets. The long-term markets include forward and futures markets, where electricity is traded up to four years in advance and as close as one month before delivery, either via power exchanges or through bilateral over-the-counter (OTC) agreements.
The short-term markets consist of the day-ahead (DA), intraday, and balancing markets. The spot market—comprising the day-ahead and intraday markets—is operated by the European Power Exchange (EPEX) and is accessible only to Balance Responsible Parties (BRPs)~\cite{35}.
BRPs trade on behalf of their customers by submitting bids and offers for next-day delivery, as well as managing their surplus or shortage in each PTU on the intraday market. 
\subsubsection{Imbalance market}
\label{2.1.2}
\begin{figure}[htbp]
\centerline{\includegraphics[width=0.9\columnwidth]{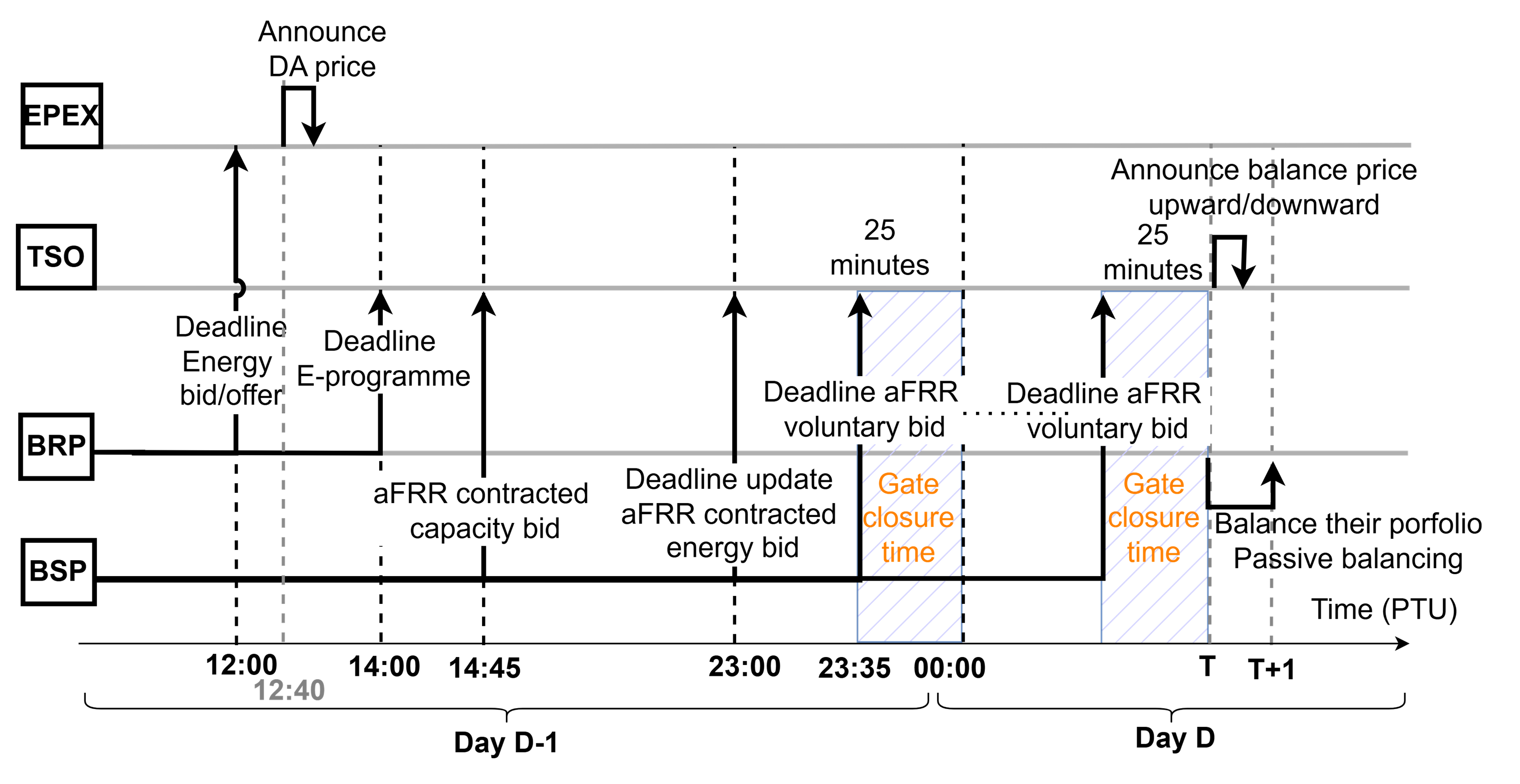}}
\caption{Market timeline.}
\label{fig:market_timeline}
\end{figure}

In real-time, TenneT - the TSO of the Netherlands - monitors the real-time balance in the country as a whole and balances the system by using balancing services from the imbalance market.
These services are provided by Balancing Service Providers (BSPs).
The balancing price is calculated using the pay-as-cleared mechanism (also known as uniform pricing or marginal pricing), whereby a bid ladder is formed based on all bids submitted by BSPs for a given time interval~\cite{Tennet_imbalance_pr}. The TSO then creates a merit order list, ranking the bids from lowest to highest price. The price of the last accepted bid is applied uniformly to all selected BSPs and is published on TenneT’s public website three minutes before real time~\cite{33}.
The imbalance price applied to BRPs is derived from the balancing price and, in most cases, is equal to it.
A summary of the timing and interaction between the system operator, BRP, and BSP is presented in Fig.~\ref{fig:market_timeline}.

\subsection{Imbalance pricing mechanism}
\begin{figure}[htbp]
\centerline{\includegraphics[width=0.9\columnwidth]{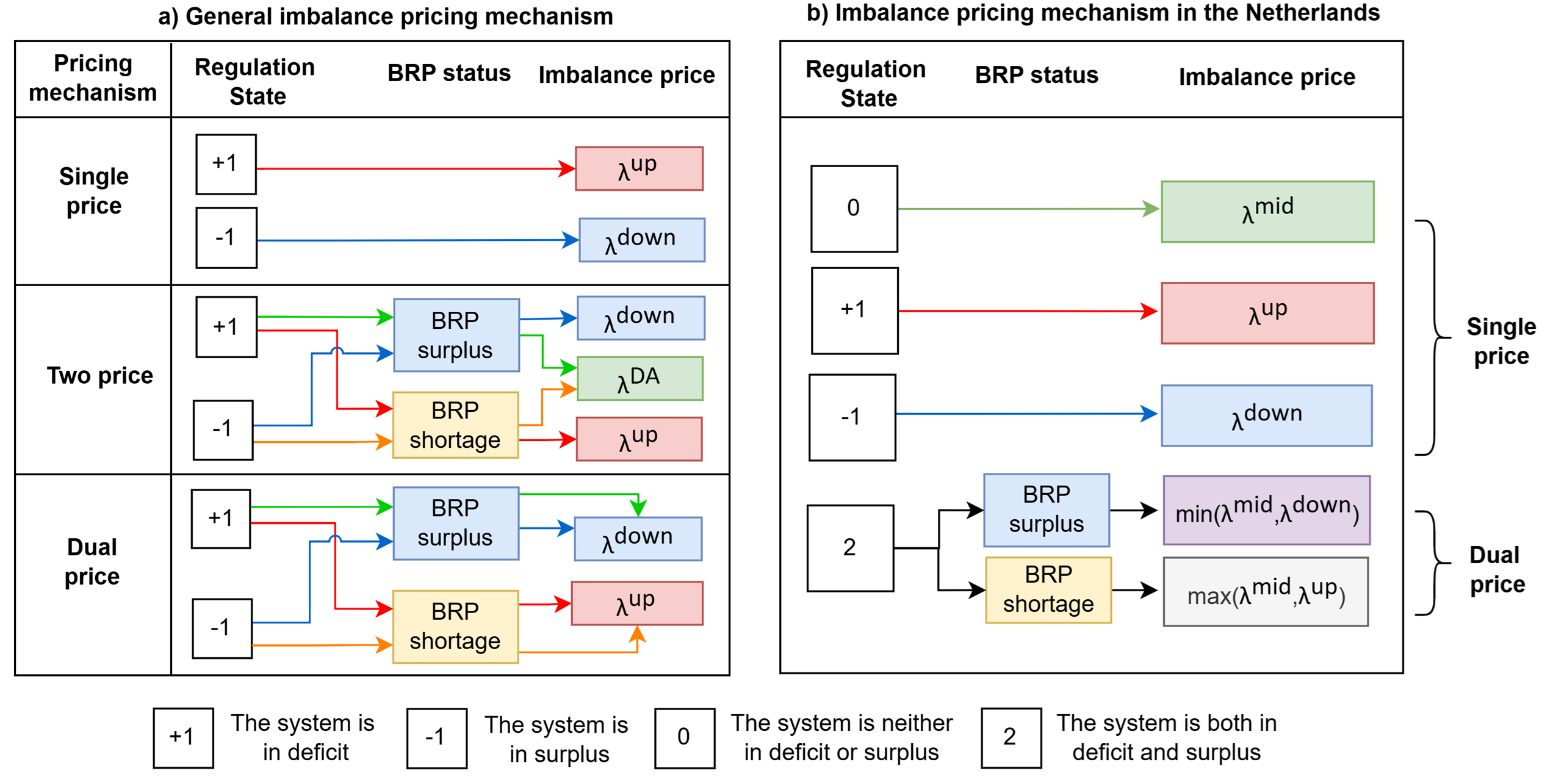}}
\caption{a) General imbalance pricing mechanisms, b) Imbalance pricing mechanism in the Netherlands.}
\label{fig:3IPMs}
\end{figure}

According to the authors in~\cite{36}, there are three main IPMs: The single-price, two-price, and dual-price mechanisms. A single price settles the BRP deviation based solely on the system direction, regardless of the imbalance directions of the BRP. 
This mechanism is set as the default across Europe by the European Union’s Agency for the Cooperation of Energy Regulators (ACER) when setting the harmonisation of the imbalance settlement protocol since 2020~\cite{acer_decision}.
Contrarily, in the two-price and dual-price mechanisms, the imbalance price is defined depending on the alignment between the BRP and the overall system imbalance direction. 
If the BRP position is opposite the system imbalance, the settlement may use the DA price, a predefined mid-price, or the opposite imbalance price. The single-price mechanism will encourage the BRP to do passive balancing, while the two-price and dual-price mechanisms almost prevent that.
The three mechanisms are described in Fig.~\ref{fig:3IPMs}a.

In the Netherlands, the TSO recognizes the positive contribution of BRPs to system balancing, as demonstrated by the various incentives provided through the IPMs. 
For this reason, the single-price is applied under three conditions: (1) when the system is short (deficit) (demand for electricity exceeds its supply), which is defined as regulation state 1; or (2) when the system is long (surplus) (supply for electricity exceeds its demand), defined as regulation state -1; and (3) when the system is neither short or long (regulation state 0), during a period in which the TSO financially settled the BRP discrepancy. 
In addition, since this region allows for passive balancing, the TSO provides incentives by publishing and continuously updating information on the activated amount of balancing energy and the corresponding price as mentioned in \ref{2.1.2}.
The dual-price approach is applied in the regulation state 2. Additionally, a mid-price - the average of the highest price for downward and the lowest price for upward regulation - is applied when the upward regulation price is lower than the
mid-price, or the downward regulation price is higher than the mid-price and in the regulation state 0.
In this study, we aim to align with the actual pricing mechanism to assess its impact and propose a suitable solution for congestion management close to real-time. Accordingly, the IPM reflects the reality while simplifying the mid-price calculation by averaging the upward and downward regulation prices.
The IPM applied in the Netherlands is shown in Fig.~\ref{fig:3IPMs}b.

\subsection{Literature review}
Considerable attention in the literature has been given to refining the balancing market design and exploring BRPs’ strategic behavior. 
In~\cite{11}, the authors propose a new market framework where any market actor is allowed to trade with an imbalance to quantify how profitable it is for the seasonal thermal energy storage system (STESS) to participate in the proposal market as well as the balance cost saving for TSO. Results show that the new market solves the strict prequalification and time frame issues with free bids, and has the potential to lower the balancing price. 
An assessment framework via market access and auction configuration was validated with the balancing market of 3 countries to determine the market variables that influence the DER integration, as introduced by~\cite{10}. The findings indicate that flexible pooling conditions, higher bidding frequency, product resolution, and authorization of non-precontracted bids can significantly ease DER integration in the market.
The impact of different IPMs on balancing market performance is assessed in~\cite{12,15} to analyze the interplay between energy and ancillary service offers, and examine the RES aggregator's motivation to bid for ancillary services. The results show that the single-price mechanism leaves room for strategic bidding by the aggregator through passive balancing, resulting in the highest profit. Additionally, there is less incentive to provide upward services than to offer downward services.

Another research direction focuses on mitigating BRP imbalance by utilizing available flexibility at residential consumers~\cite{7,22} and online scheduling for storage~\cite{7}, through Virtual Power Plan support~\cite {16}, cooperation among BRP members~\cite {13}, or by separating the imbalance price by direction~\cite{17}. The authors in~\cite{23} demonstrate that by applying predictive real-time dispatch and changing PTU length, more balancing effort is allocated to BRPs than to TSO, then helps reduce the system imbalance.
By considering the BRP position, system sign, and impact of the BRP member on the imbalance of BRP, an improvement in the cost-revenue allocation at the level of BRP is introduced in~\cite{13}. This allocation ensures a transparent and equitable setting of the imbalance, BRP members benefit from cooperating, and BRP minimizes the imbalance cost.
However, there is limited literature focusing on the impact of the BRP action and the pricing mechanism on the distribution grid.
In~\cite{14,6}, the BRP impact on the grid and the economic benefit are explored in the context of an energy storage system (ESS) utilization~\cite{14}, or enhancing the BRP and DSO coordination~\cite{6}, while neglating the imbalance pricing mechanism.
The literature is compared by different criteria in Table~\ref{tab0}.

\begin{table}[htbp]
\centering
\caption{Comparison of the existing research with this work}
\fontsize{8.5}{12}
\selectfont
\begin{tabular}{|m{3.5cm}|m{1.8cm}|m{3.5cm}|m{1.5cm}|m{1.5cm}|}
\hline
\multirow{2}{*}
 
\textbf{BRP model} & 
\textbf{Uncertainty} & 
\textbf{Pricing mechanism} & 
\textbf{Grid constraint} &
\textbf{Ref}
\\
\hline
    Deterministic  &\text{No} &  No & Yes &~\cite{6} \\
    optimization &\text{No} & Single pricing scheme &  \text{No} &~\cite{7}\\
\hline
     & Yes & Single/Two/Dual-price &  \text{No} &~\cite{1} \\
     Agent-based modelling & \text{Yes} & No &  \text{No} &~\cite{11} \\
     & \text{No} & Single and Two-price &  \text{No} &~\cite{40} \\
\hline
    Two-dimension quantile forecast function & \text{Yes} & Dual-price with penalties &  \text{No} &~\cite{3}\\
\hline
    Probabilistic-based method&\text{Yes} & Non linearity and Penalty &  \text{No} &~\cite{9}\\
\hline
    Hierachical model predictive control &\text{Yes} & No &  \text{No} &~\cite{42} \\
\hline
    Low cost generators and high cost generators are modeled & \text{Yes} & Single, Typical-Dual, Renewable-Dual, and the proposed optimal pricing mechanism &  \text{No} &~\cite{43}\\
\hline   
    Stochastic risk-constrained optimization & \text{Yes} & Single/Two/Dual-price &  \text{No} &~\cite{12}\\
\hline
    Stochastic model predictive control & \text{Yes} & No &  \text{Yes} &~\cite{41}\\
\hline   
      & \text{} & One price average for the whole day &  \text{No} &~\cite{16}\\   
    Stochastic optimization & \text{Yes} & Proposed piece-wise linear penalty pricing scheme &  \text{No} &~\cite{2}\\
     & \text{} & Single and Two-price &  \text{No} &~\cite{39} \\
\hline     
     &\text{} & No &  \text{No} &~\cite{8,13,20}  \\  
     & \text{} & Single , dual and the proposed separate mechanism &  \text{No} &~\cite{17}\\
   No mentioned  & \text{No} & Yes &  \text{No} &~\cite{18,38} \\
     & \text{} & Single/Dual and Hybrid imbalance pricing &  \text{No} &~\cite{37} \\
     & \text{} & Nodal Single-price &  \text{Yes} &~\cite{29} \\
\hline
    \textbf{Stochastic optimization} & \textbf{Yes} & \textbf{Single/Two/Dual-price} &  \textbf{Yes} & \textbf{This work}  \\
\hline
\end{tabular}
\label{tab0}
\end{table}

The above research provides discussions on (1) the loophole in the market design, such as inconsistent pricing mechanisms, price predictability, and passive balancing incentive, (2) introduces a new balancing market design that allows a higher DER share, and (3) introduces different methods to mitigate BRP deviation and the impact on the distribution grid. 
There is a gap between the system-level design of the balancing market and imbalance pricing mechanisms, and local-level challenges such as the impact on the distribution network and the evaluation of information flow among actors.
To the best of our knowledge, this is the first attempt to combine the study of different IPMs with consideration of congestion in the distribution network and the impact of BRPs' response to such incentives. Therefore, this paper presents the following contributions:
\begin{itemize}
\item Proposing a congestion-dependent imbalance pricing mechanism for preventing congestion caused by the simultaneous reaction of BRP to the attractive high or low imbalance price:  The proposed pricing mechanism requires enhanced coordination between DSOs and BRPs and localizes imbalance pricing by separating congested and non-congested areas.
\item Presenting a comprehensive two-stage stochastic model for BRP decision making under different IPMs, while considering both DA and imbalance price uncertainty.
\item Providing proof of concept via a case study in which the pricing mechanism and the BRPs' response are demonstrated in a real Dutch distribution network.
\end{itemize}
The remainder of the paper is organized as follows. Section \ref{propose_method} discusses the proposed pricing mechanism to solve the congestion issue caused by BRP actions. Section \ref{BRP_model} introduces the mathematical model of the two-stage BRP model with different pricing mechanisms. Simulation and the case study via different cases are demonstrated in section \ref{case_study}. Results will be presented and evaluated in Section \ref{simulation}. Finally, conclusions and discussion on future work are provided in the last section.

\section{Proposed congestion-dependent imbalance pricing mechanism }
\label{propose_method}
\begin{figure}[htbp]
\centerline{\includegraphics[width=0.75\columnwidth]{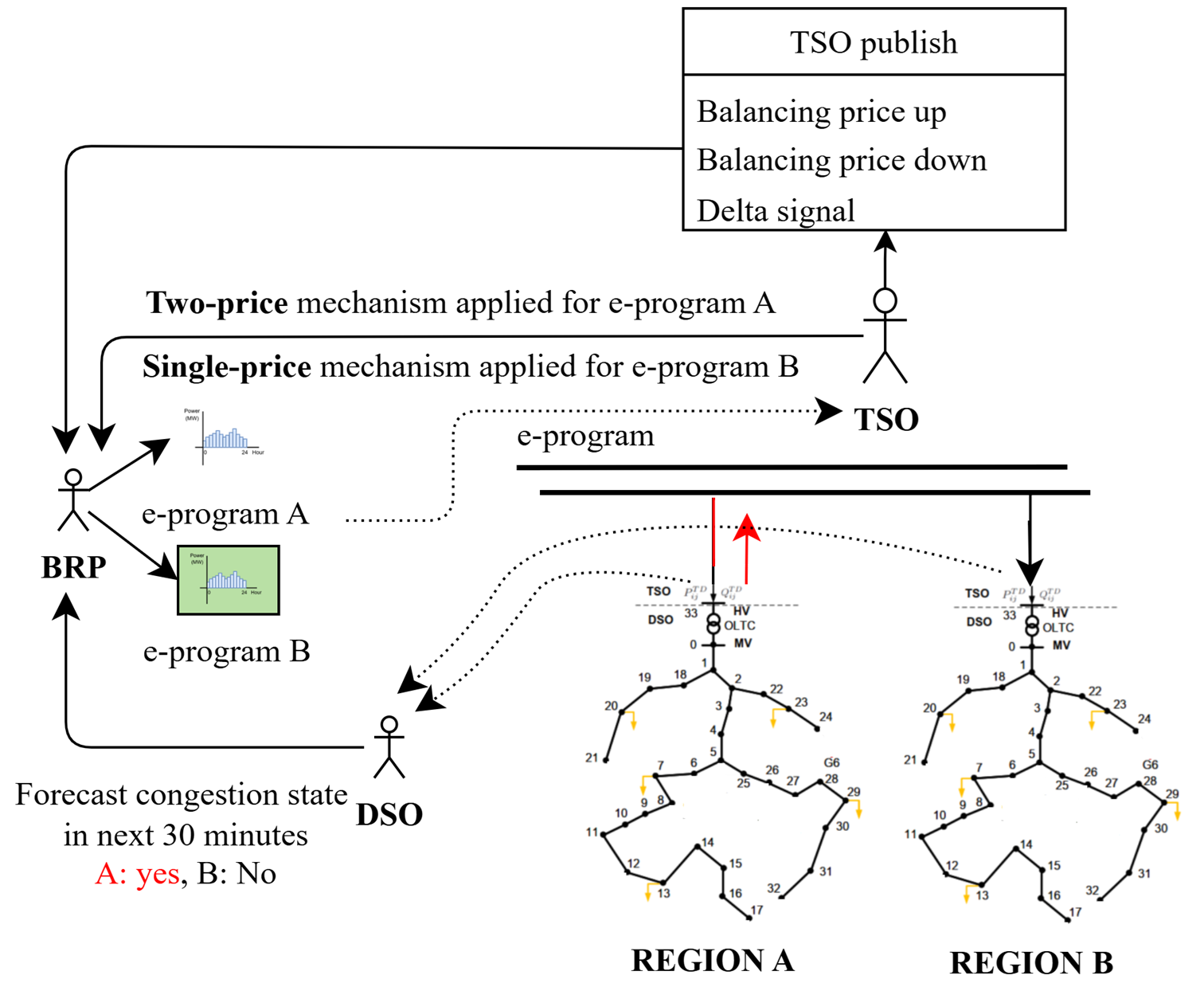}}
\caption{Proposed mechanism.}
\label{fig: mechanism}
\end{figure}

As described in the literature review section, the single price mechanism helps BRPs achieve the lowest imbalance cost, and serves as an effective incentive for them to support system balancing. Contrarily, the dual price and two price mechanisms offer limited economic benefit and are not attractive enough to encourage the BRP to engage in passive balancing, as they often result in higher imbalance costs. One hypothesis is that the incentive from the single price mechanism may trigger simultaneous reactions from many BRPs. If these passive balancing actions are spatially concentrated, they can adversely affect the distribution grid, resulting in congestion in that area, which will be reproduced in the simulation result.
Meanwhile, the imbalance deals with the physical delivery of energy, further reinforcing the argument that grid constraints should not be ignored~\cite{29}.
Therefore, it is necessary to strike a balance between maintaining the financial incentives and mitigating the negative impact on the distribution grid. In this study, we propose a congestion-dependent imbalance pricing approach that:
\begin{itemize}
    \item Enhances the coordination between DSOs and BRPs: The DSOs will forecast the demand and generation for the next 2 PTUs for a specific region, before the closing time for TSO to activate the balancing service. Then, they inform the congestion state in that region to the relevant BRPs and the TSO.
    \item Spatially separating the imbalance pricing mechanism: Congested areas will apply the two-price or dual price mechanism, with less incentive for BRPs to do passive balancing. Contrarily, non-congested areas will maintain the single price mechanism.
    \item Settlement: TSO settles the BRP's deviation based on the localized e-program by the associated pricing mechanism.
\end{itemize}
The interaction between the actors is explained in detail in Fig.~\ref{fig: mechanism}.
Table~\ref{tab2} describes the difference in pricing mechanisms in a congested area and a non-congested area.

\begin{table}[htbp]
\centering
\caption{Differences in pricing mechanisms in congested and non-congested areas}
\fontsize{8.5}{12}\selectfont
\begin{tabular}{|m{3cm}|m{2.5cm}|m{2.5cm}|m{3.5cm}|}
\hline
\multirow{2}{*}

\textbf{} & 
\textbf{System state} & 
\textbf{BRP state } &
\textbf{Pricing mechanism }
\\
\hline
        \textbf{} & Long (-1) & Shortage & \textbf{Single price} \\
        \textbf{Without congestion} & Long (-1) & Surplus & Single price \\
        \textbf{} & Short (+1) & Shortage & Single price \\
        \textbf{} & Short (+1) & Surplus & \textbf{Single price} \\
\hline
         \textbf{} & Long (-1) & Shortage & \textbf{Two price/Dual price} \\
        \textbf{With congestion} & Long (-1) & Surplus & Single price \\
        \textbf{} & Short (+1) & Shortage & Single price \\
        \textbf{} & Short (+1) & Surplus & \textbf{Two price/Dual price} \\
\hline
\end{tabular}
\label{tab2}
\end{table}

We now illustrate the proposed pricing mechanism by means of an example. Suppose region A is experiencing congestion at its connection point to the transmission system, with power flowing from the DN to the TN. Meanwhile, in region B, power flows from the TN to the DN without any congestion. The TSO detects a system-wide imbalance resulting in a deficit. In response, the TSO requests upward regulation, and the upward regulation prices are published, along with the activated volumes for the previous three minutes. This creates an incentive for BRPs in both regions A and B to either generate more or consume less to support system balance. In this case, BRP represents groups of EV fleets that prefer to consume less.
If the BRPs in region B reduce consumption, it helps alleviate the system imbalance. However, if the BRPs in region A reduce consumption, it increases the reverse flow from the DN to the TN via an already congested connection point, worsening the situation. 
If no action is taken, this situation will persist and the ongoing congestion can negatively impact the lifetime of components at the substation or lead to a brownout in the region.
In case DSO predicts this congestion situation, they inform the BRPs in region A ``Yes'' flag, which means congestion will happen in this region.
If a BRP in this region does passive balancing, its deviation is settled based on a specific imbalance pricing mechanism—either the two-price or dual-price mechanism, referred to as IPM A. The BRP recalculates its optimal deviation from the contracted value by incorporating IPM A into its decision-making.
After the ISP concludes, the TSO settles BRPs in region A using IPM A, while BRPs in region B continue to be settled under the single pricing mechanism, referred to as IPM B. In this way, BRPs in the congested area have an incentive to stay on their schedule and avoid actions that increase the congestion.

\section{BRP MODEL}
\label{BRP_model}
\begin{figure}[t]
\centerline{\includegraphics[width=0.8\columnwidth]{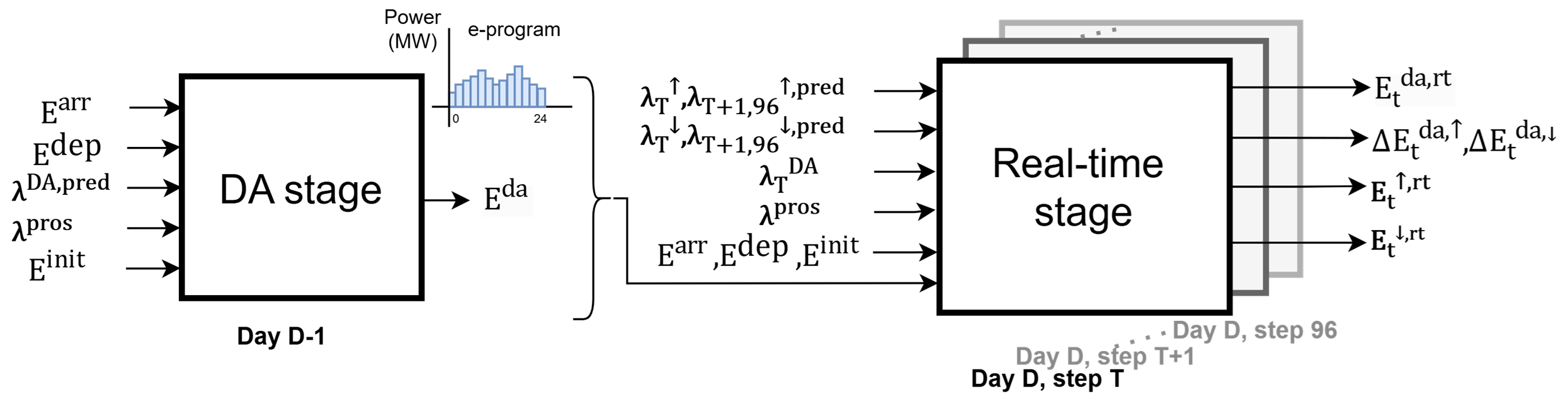}}
\caption{BRP 2-stage models.}
\label{fig:brp_model}
\end{figure}

To represent the passive balancing behavior of BRPs under different IPMs, this study proposes a two-stage stochastic model for the BRP managing  Electric Vehicle (EV) fleets. Each BRP manages several groups of 100 EVs, which are modeled as a virtual battery (VB). The VB model is based on~\cite{24}, and the EV data is generated based on the general mobility data of Dutch citizens from~\cite{25}. BRPs consider uncertainties of the DA price, imbalance price for upward and downward regulation, and EV data as input. The inputs and outputs of each stage are depicted in Fig.~\ref{fig:brp_model}. Besides, the term PTU will be used in the DA stage of the model. After the event, the term ISP is used to describe the settlement period, the value is the same as PTU in the case of the Netherlands.

Several assumptions are made:
\begin{itemize}
    \item BRPs have perfect information about the EV session of their customer, and the EV owner follows their contract.
    \item The BRP e-program at the output of the DA stage is the final version submitted to the TSO.
    \item The imbalance price at the current step is the actual price, the price for the next step until the end of the day is based on the forecast scenario.
    \item EV can only be charged at home, and Vehicle-to-grid (V2G) is not allowed.
    \item The regulation state of the system is assumed to be correct for actual operation.
\end{itemize}

\subsection{Virtual Battery}\label{AA2}
The required information for EV comprises the initial state of charge (SOC), arrival time, departure time, and energy required for the trip. The virtual battery (VB) upper and lower charging paths are determined by a bottom-up approach, starting from the charging path of each individual EV via the as soon as possible (ASAP) and as late as possible (ALAP) principle, explained in detail in~\cite{24,Bart_paper} and listed in the Appendix section.

\subsection{DA stage with single price mechanism}\label{AA3}
In the first stage, based on the forecast DA price and the trip session from their customer, BRP runs the optimization model to get the optimal bidding schedule, which determines the energy they will buy from the DA market $E^{da}_{t}$ to charge the EV battery at a specific time the next day. The aggregation of all bids/offers by time - the commercial trade schedule - is then submitted to TSO and is used as input to the second stage. $S_{da}$ presents a set of DA price scenarios, where each scenario occurs with an equal probability $\pi_{s_{da}}$.
The objective function of BRP is to maximize its benefit from buying energy in the DA market with price $\lambda_{s_{da},t}$ and selling to its customers at a fixed higher price $\lambda^{pros}$ by Eq.~\ref{eq:obj}, the relevant constraints are presented in the Appendix.
\begin{align}
\max \sum_{s_{da}}^{S_{da}}\pi_{s_{da}}\sum_{t}^{T} (E^{da}_{t} (\lambda_{s_{da},t} - \lambda^{pros})) \label{eq:obj}
\end{align}

\subsection{Real-time stage with single price mechanism}\label{AA1}
In the real-time stage, BRPs have the opportunity to either correct their imbalance or deliberately create deviation to support system balancing. This is made possible by the information about the imbalance price published by TSO shortly before real-time. 
Input of this stage includes the actual imbalance price at the current step, forecasted imbalance price from the next step until the end of the day, the known DA price, energy purchased in the DA market, and EV data. In each ISP, the BRP performs an optimization problem to get the actual volume they should buy. The deviation from the energy bought from DA market $E^{da}_t$ is represented as a delta surplus $\Delta E^{\uparrow}_{t}$ or shortage $\Delta E^{\downarrow}_{t}$. The objective is to minimize the actual imbalance cost, or to maximize the potential gain from passive balancing as formulated in Eq.~\ref{eq:EV_rt}. This model also takes into account the future information of the price and EV session by rolling horizon with a 15-minute resolution.
Given the stochastic nature of the imbalance price, a random error with a standard deviation of 50\% is added to the actual price to generate 5 scenarios for upward regulation price and 5 scenarios for downward regulation price. This results in a total of 25 real-time scenarios, the number of price scenarios selected is similar to~\cite{12}.
The associated constraints are described in Eq.~\ref{eq:gh_constraint} and detailed in the Appendix.
\begin{align}
\max \sum_{s_{rt}}^{S_{rt}} \pi_{s_{rt}}\sum_{t=1}^{T}  
    (\Delta E^{\uparrow}_{t} +  \Delta E^{\downarrow}_{t})
    (\lambda^{\uparrow}_{s_{rt},t}g_t + 
     &\lambda^{\downarrow}_{s_{rt},t}h_t)/2
    + E^{da}_t (\lambda^{da}_t - \lambda^{pros}) \label{eq:EV_rt}
\end{align}

In which, $\lambda^{\uparrow}_{s_{rt},t}$ and $\lambda^{\downarrow}_{s_{rt},t}$ represent the forecasted imbalance price for upward and downward regulation at time step t, in real-time scenario. $\lambda^{da}_t$ is the actual value of DA price.  $S_{rt}$ presents a set of real-time price scenarios, where each scenario occurs with an equal probability $\pi_{s_{rt}}$
The objective function is formed in combination with the binary variables $g_t, h_t$ to ensure only the upward/downward regulation price or mid-price is applied when the regulation state equals 1/-1 or 0/2, respectively. 
\begin{align}
g_t, h_t =
\begin{cases}
2,\, 0 & \text{if } \text{state} = 1 \\
0,\, 2 & \text{if } \text{state} = -1 \\
1,\, 1 & \text{otherwise}
\end{cases} \quad\forall t
\label{eq:gh_constraint}
\end{align}

\subsection{Real-time stage with two-price mechanism}
With the two-price mechanism, the objective function of BRPs is modified as follows:
\begin{align}
\max \sum_{s_{rt}}^{S_{rt}} &\pi_{s_{rt}}\sum_{t=1}^{T}  
    (E^{da}_t (\lambda^{da}_t - \lambda^{pros}) 
    +(\Delta E^{\downarrow}_{t}\lambda^{\uparrow}_{s_{rt},t} + \Delta E^{\uparrow}_{t}\lambda^{da})g_t +
    \notag\\
    &(\Delta E^{\downarrow}_{t}\lambda^{da} +
    \Delta E^{\uparrow}_{t}\lambda^{\downarrow}_{s_{rt},t})h_t
    +(\Delta E^{\uparrow}_{t} + \Delta E^{\downarrow}_{t})(\lambda^{\uparrow}_{s_{rt},t} +\lambda^{\downarrow}_{s_{rt},t})k_t/2) \label{eq:EV_rt_two}
\end{align}
Subject to:
\begin{align}
g, h, k =
    \begin{cases}
1,0,0 & \text{if } \text{state} = 1 \\
0,1,0 & \text{if } \text{state} = -1 \\
0,0,1 & \text{otherwise}
\end{cases}
\end{align}
The binary variables $g_t, h_t, k_t$ ensure that the upward regulation price is applied for the BRP with surplus, and the DA price applies for the BRP with shortage when the regulation state equals 1. Contrarily, the downward regulation price is applied for the BRP with a shortage, and the DA price applies for the BRP with a surplus when the regulation state is -1. The mid-price is used in the remaining cases.
\subsection{Real-time stage with dual price mechanism}
Similarly, the objective function of the BRP in this case is adjusted as follows:
\begin{align}
\max \sum_{s_{rt}}^{S_{rt}} \pi_{s_{rt}}\sum_{t=1}^{T}  
    (E^{da}_t (\lambda^{da}_t - \lambda^{pros})
    +\Delta E^{\downarrow}_{t}\lambda^{\uparrow}_{s_{rt},t} + \Delta E^{\uparrow}_{t}\lambda^{\downarrow}_{s_{rt},t}) 
     \label{eq:EV_rt_dual}
\end{align}
With the dual-price mechanism, the deviation of BRP is settled based on their position, regardless of the system direction. The BRPs that cause shortages are always paid a penalty to the TSO with the upward regulation price, and the BRPs that cause surpluses are always paid with the downward price. The other constraints are used as in the two previous cases.
The optimization problems are mixed-integer linear programming (MILP) and are solved by the Gurobi solver in Pyomo.

\section{Case study}
\label{case_study}
\subsection{Network description}
The BRP's behavior in the different IPMs, as well as its impact on the DNs, are investigated by using the real distribution network at a substation in the northern Netherlands. This network is operated at the voltage level of 10.5 kV, sources by 4 step-down two-winding transformers from the external grid at 25 kV. 
The total installed capacity is 66.06 MW, with a total spinning reserve is 11.36 MW. The PV penetration rate is 12.85\% (8.5 MW) from 16 installation points.
The BRPs VB connected to nodes 145, 146, 147, 176, and 182 are considered as region A. Then, the BRPs VB connected to nodes 42, 103, 111, 113, and 142 are region B.
Each EV is assigned randomly a maximum capacity of 50 or 75 kWh, and a maximum charging rate of 3.7 kW or 11 kW, respectively. 

\subsection{Simulation Cases}
Pricing data from one year, from the second half of 2023 to the first half of 2024, collected from the ENTSOe-transparent platform, is used to validate the simulation with 3 different pricing mechanisms in global and local cases:

\begin{itemize}
    \item Single price mechanism (SP)
    \item Two-price mechanism  (TP)
    \item Dual price mechanism (DP)
\end{itemize}

The global cases mean the pricing mechanism is applied uniformly across the entire system (both regions A and B). In contrast, the local cases implement the two-price or dual-price mechanism specifically in the congested area, while maintaining the single price mechanism in the non-congested area.
The simulation result is evaluated by the BRP benefit, and the congestion situation is quantified by the loading of the line at the connection point.

\section{Simulation result}
\label{simulation}
\subsection{BRP behavior via different IPMs} 
A BRP receives money when they have a surplus and pays when they shortage. Whether there is a benefit depends on the price difference between the regulation price and the wholesale (DA) price. A BRP has a benefit from passive balancing when its position is against the system's direction.

\subsubsection{Single price mechanism}
\begin{figure*} 
\subfloat[Single-price mechanism]{\centerline{\includegraphics[width = 1\columnwidth]{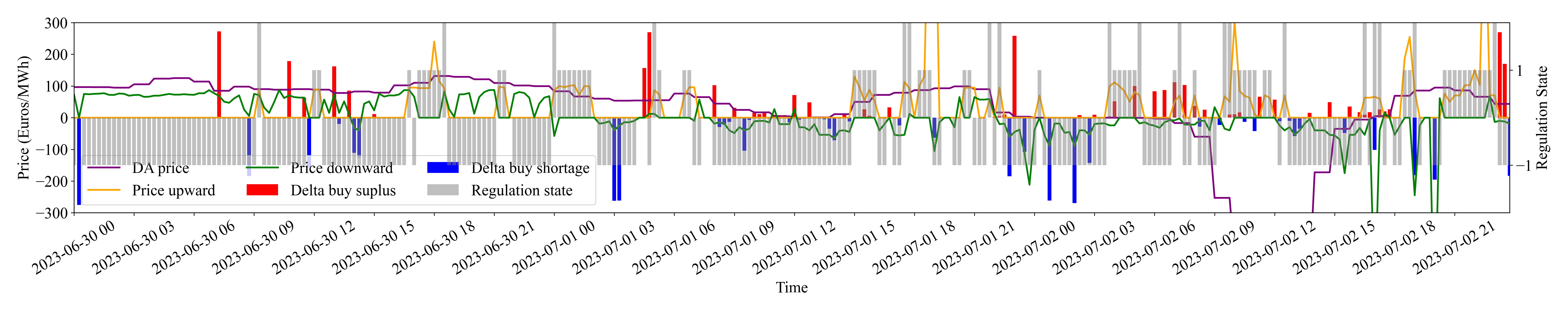}}}\\ 
\subfloat[Two-price mechanism]{\centerline{\includegraphics[width = 1\columnwidth]{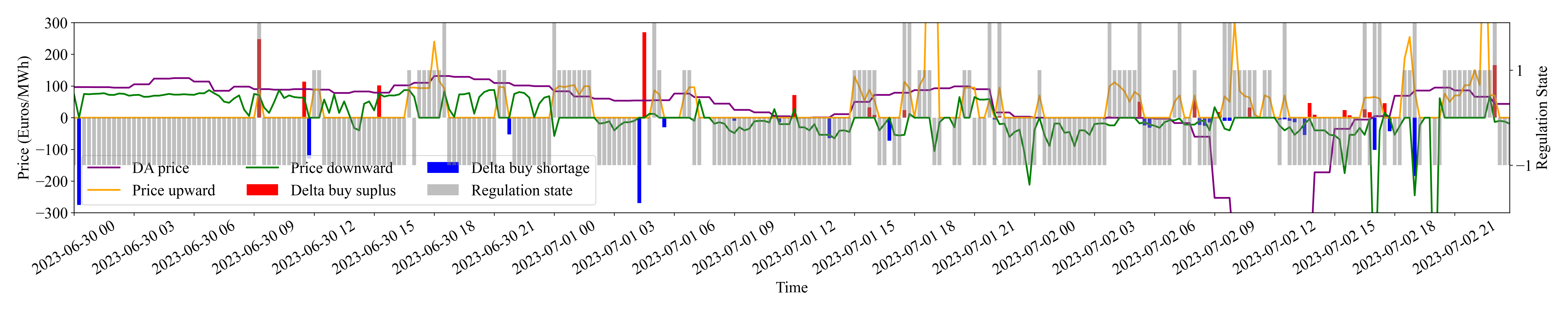}}}\\ 
\subfloat[Dual-price mechanism]{\centerline{\includegraphics[width = 1\columnwidth]{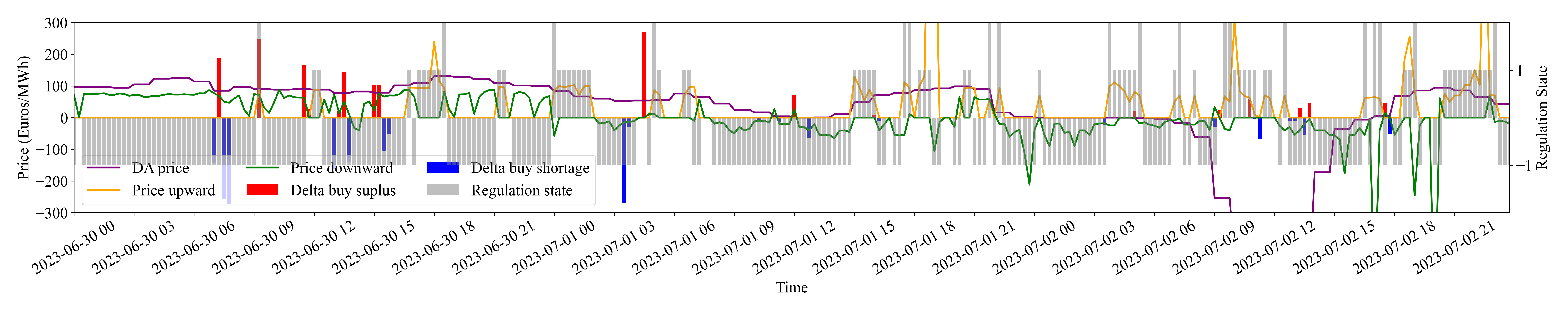}}} 
\caption{BRP surplus and shortage in real-time}
\label{fig: rt_sing}
\end{figure*}

\begin{figure*}[htbp]
\subfloat[Single-price mechanism]{\centerline{\includegraphics[width = 0.75\columnwidth]{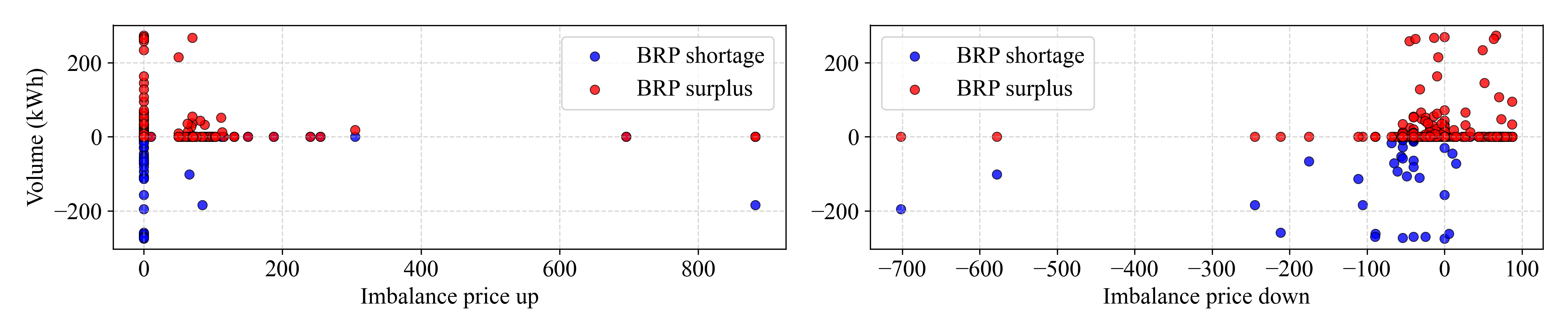}}}\\
\subfloat[Two-price mechanism]{\centerline{\includegraphics[width = 0.75\columnwidth]{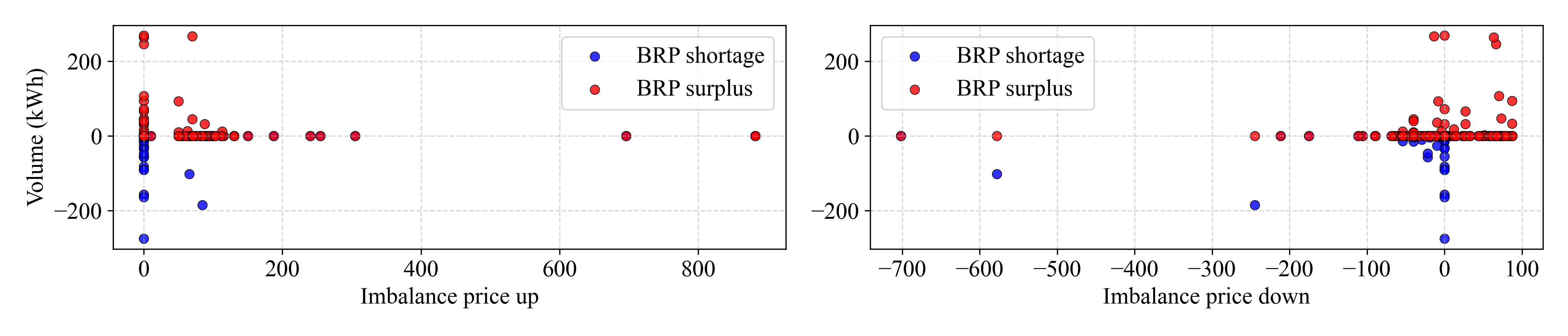}}}\\
\subfloat[Dual-price mechanism]{\centerline{\includegraphics[width = 0.75\columnwidth]{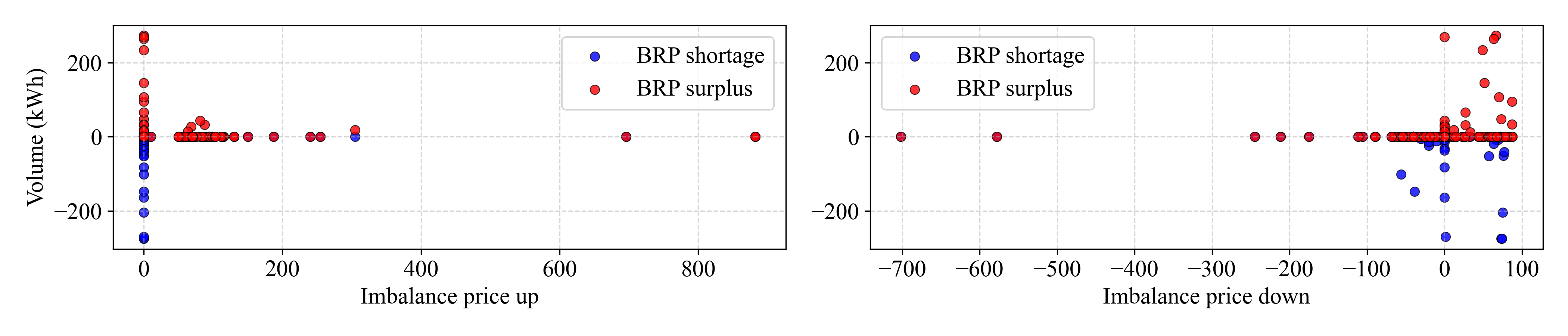}}}
\caption{Distribution of surplus and shortage volume by the upward (left), and downward regulation price (right)}
\label{fig: dis_sing}
\end{figure*}

 In the single-pricing mechanism, there are four options when the direction of payment from TSO to BRP:
 \begin{enumerate}
     \item BRP shortage, the system is long (regulation state is +1), and the downward-regulation price is negative,
     \item BRP surplus, the system is short (regulation state is -1), and the upward-regulation price is positive,
     \item BRP surplus, the system is long (regulation state is +1), and the downward-regulation price is positive,
     \item BRP shortage, the system is short (regulation state is -1), and the upward-regulation price is negative.
 \end{enumerate}
The simulation result of 3 days in the summer of 2023 is presented in Fig.~\ref{fig: rt_sing}a and  \ref{fig: dis_sing}a. 
In which, Fig.~\ref{fig: rt_sing}a shows the price in (\euro/MWh) and energy volumes in kWh over the simulation time horizon. Fig.~\ref{fig: dis_sing}a presents the distribution of BRP shortages and surpluses under upward and downward regulation prices (\euro/MWh).
Options (1) and (2) bring the most benefit for BRP due to the occurrence and the high absolute value of the imbalance price, represented in the third quadrant of the right chart and the first quadrant of the left chart in Fig.~\ref{fig: dis_sing}a. The red bars in Fig.~\ref{fig: rt_sing}a or the red dots in Fig.~\ref{fig: dis_sing}a, represent the surplus volume. Similarly, the blue bars and the blue dots show the shortage volume.
The gray bar represents the regulation state of the system. 
Option (3) happens due to the must-run conditions of generators, resulting in a positive price for downward regulation.
This case is shown in the first quadrant of the right chart in Fig.~\ref{fig: dis_sing}a or the red bar when the gray bar is showing -1 and the downward regulation price (green line) is positive.
Moreover, it is rarely the case that the upward-regulation price is negative. Therefore, option (4) does not occur in this simulation time period, and no dots are distributed in the second and third quadrants of the left chart. 

\subsubsection{Two price mechanism}
Differently, the two-price mechanism does not encourage strategic behavior from the BRP. Therefore, the system direction is considered by applying the day-ahead price to the case when the BRP direction is opposite to the system direction. BRPs receive money mostly when they surplus, and the downward price or the DA price is high. The red dots concentrate more on the first quadrant on both the left and right charts of Fig.~\ref{fig: dis_sing}b or red bars in Fig.~\ref{fig: rt_sing}b. 
However, there is no shortage when the downward price is high, due to the settled price in this case is the DA price, which brings no benefit to the BRP, and they try to stick with the schedule. 
Several outliers were detected in the fourth-quadrant of the left chart and the third-quadrant in the right chart belong to (1) the states 0 and 2 when the mid price is applied or (2) when the system is in deficit and the DA price is negative, then this shortage brings profit to BRPs.
\subsubsection{Dual price mechanism}
The dual-pricing mechanism does not consider the system direction. The BRPs have a benefit mostly when they surplus, and the downward regulation price is high; the red dots lie in the first quadrant on the right of Fig.~\ref{fig: dis_sing}c or red bars in Fig.~\ref{fig: rt_sing}c. Since there is less profit to be gained from arbitrage, the frequency of deviation is significantly reduced compared to the single-pricing case.
Interestingly, some surplus still happens when the upward regulation price is high (first quadrant in the left chart), and some shortage occurs when the downward regulation price is high (fourth quadrant in the right chart). This is because when the upward price is high, the downward price is almost equal to zero, which is the price that the TSO settles with the BRP surplus. Similarly, when the downward price is high, the upward price - the price that the TSO settles with the BRP shortage - is zero in most cases, which leads to no fee for charging their EV, then they can consume less when the downward price is high to reduce charging cost.

\subsection{Congestion issues with different IPMs}
\begin{figure*}[htbp]
\subfloat[Single-price mechanism]{\centerline{\includegraphics[width = 1\columnwidth]{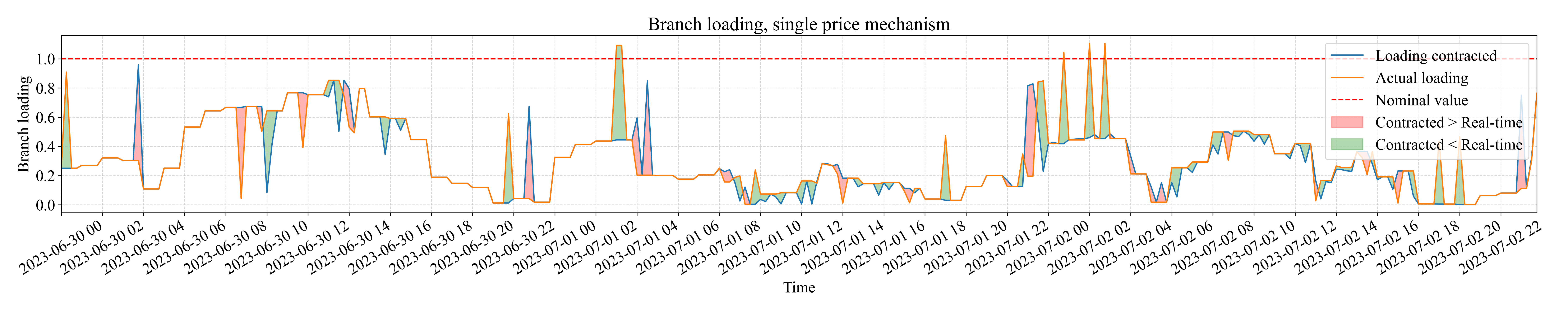}}}\\
\subfloat[Two-price mechanism]{\centerline{\includegraphics[width = 1\columnwidth]{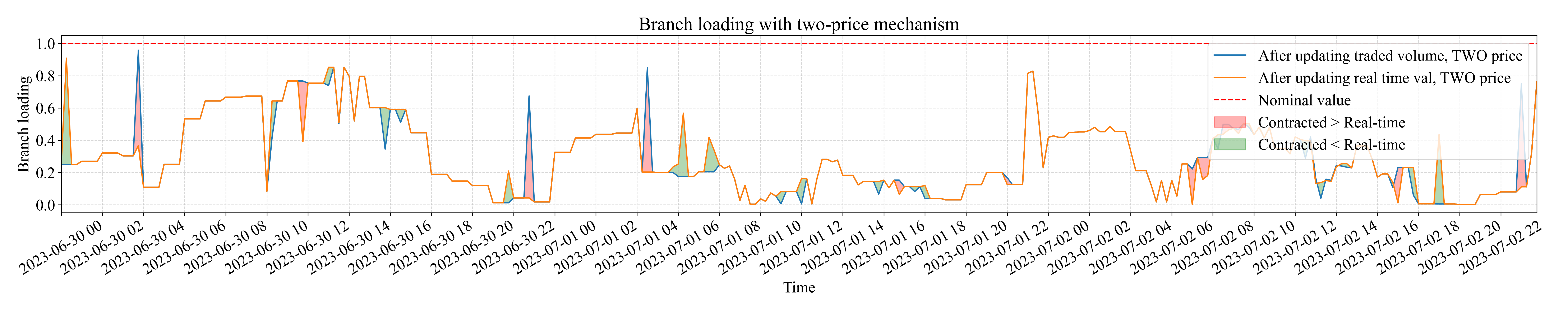}}}\\
\subfloat[Dual-price mechanism]{\centerline{\includegraphics[width = 1\columnwidth]{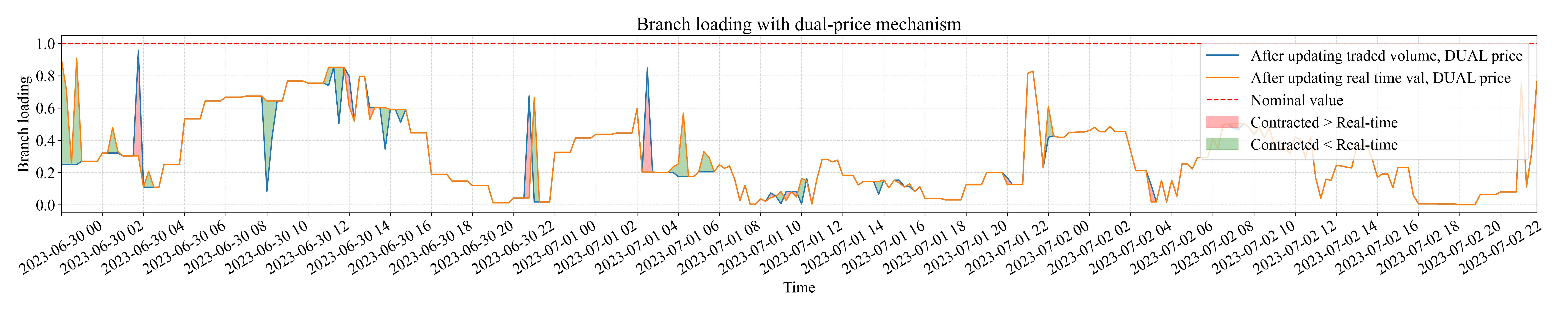}}}\\
\subfloat[Proposed imbalance pricing mechanism]{\centerline{\includegraphics[width = 1\columnwidth]{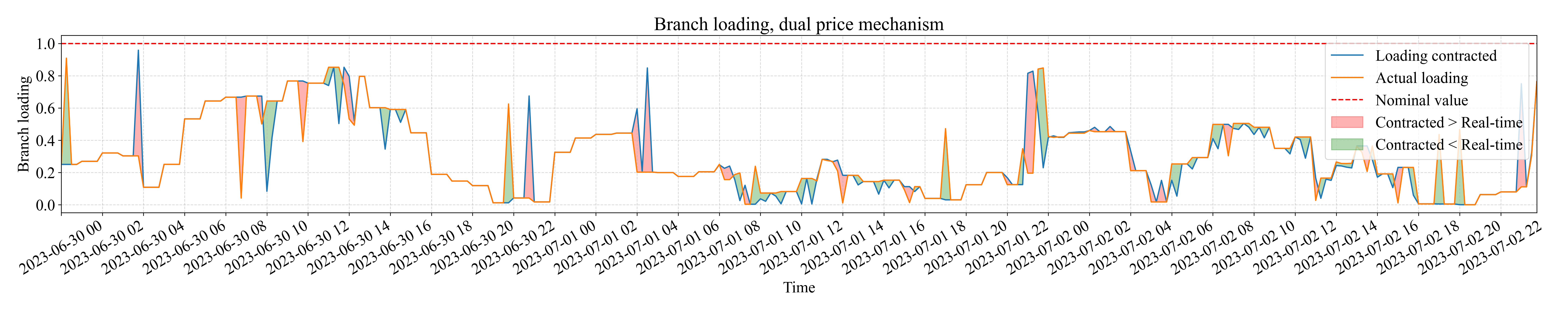}}}\\
\caption{Loading at the observed line in different IPMs}
\label{fig: load_1w_single}
\end{figure*}

While such systems offer significant benefits for the market parties, they also come with risks associated with their grid. In market designs that allow and incentivize passive balancing by the BRPs, these parties can profit, and system operators may reduce the need for costly balancing services. However, when many BRPs respond simultaneously, it can create unintended impacts on the grid. This issue is illustrated in Fig.~\ref{fig: load_1w_single}a.
The orange line shows the actual loading while the blue line shows the loading based on the schedule value. Overloading occurs when the system is long (-1), and the downward price is negative with a high absolute value. BRPs are encouraged to shortage (inject less or consume more than traded in the DA market), as presented by the blue bars in Fig.~\ref{fig: rt_sing}a or the third quadrant of the right chart in Fig.~\ref{fig: dis_sing}a. However, multiple BRP react at the same time, to the same price, causing the unexpected peak flow at ISPs 108,109, 195, 200, and 203 (around 1 am and 2 am) when the orange line crashes the dashed red line, which presents the loading limit. 
On the contrary, at the same steps, if the system only applies the dual or two price approach, no overload occurs, as shown in Fig.~\ref{fig: load_1w_single}b and Fig.~\ref{fig: load_1w_single}c. This can be explained by the lower shortage volume and less frequent occurrence of surplus/shortage in Fig.~\ref{fig: rt_sing}b and \ref{fig: rt_sing}c. 

\subsection{Congestion issue mitigation with the proposed IPM.}
This apparent behavioral difference is the basis for the new pricing method proposed earlier in section \ref{propose_method}.
The simulation result of the proposed approach is shown in Fig.~\ref{fig: load_1w_single}d.

With this approach, only BRPs in the congested area will be exposed to the new pricing mechanism when congestion is forecasted.
Therefore, the load at the time when there is no overload remains the same as before. All overloads are now resolved because the BRP adheres to the scheduled value. Another remark is that the dual-price and two-price mechanisms in this case give similar results, because there is no attraction for them to shortage in the congested region during these ISPs.

In addition to managing congestion, this approach maintains the benefit for the BRPs in non-congested areas. Table~\ref{tab3} presents the total BRPs benefit under different IPMs in the three selected days, both with and without considering congestion. When congestion is not considered, either the single-price or dual-price or two-price is applied uniformly across the entire system at all ISPs. In which, the single-price mechanism yields the highest benefit, consistent with the findings in~\cite{15}. While the dual and two-price mechanisms result in a reduction of BRPs' benefits, approximately half of that under the previous one. In case congestion is considered, in the global cases, during the congestion periods, both regions A and B will apply the dual-price or two-price mechanism. In contrast, under the local cases, only BRPs located within the congested area (A) are subject to the new pricing mechanism. There is no difference in BRP benefits between global and local implementations when only the single-price mechanism is applied, as shown in the first row. Moreover, for both the global and local cases, combining the single-price mechanism with either the two-price or dual-price mechanism yields identical outcomes, since the two-price and dual-price mechanisms are activated only during periods of congestion. However, a distinct difference emerges between local and global cases. 
A reduction of about 16\% in BRP's profit when the new pricing mechanism is applied system-wide, while only 8\% decrease is incurred if the new approach is applied only in the congested area. BRPs outside the congested area in the local setting can still engage in passive balancing and contribute to system balancing, thus gaining additional advantages.
Accordingly, this calculation proves that applying locally the new price mechanism helps maintain the incentive for the BRP in the non-congested area, which explains the higher benefit than applying globally.
This deserves consideration if the system operator takes into account the opportunity cost they have to pay for solving congestion or for compensating the component degradation.

\begin{table}[htbp]
\centering
\caption{BRP benefit (\euro) in different cases through single-price, two-price, and dual-price mechanisms}
\fontsize{8}{12}\selectfont
\renewcommand{\arraystretch}{1}
\begin{tabular}{|>{\centering\arraybackslash}m{2cm}|>{\centering\arraybackslash}m{1cm}|
                >{\arraybackslash}m{2.4cm}|>{\centering\arraybackslash}m{1cm}|
                >{\arraybackslash}m{2.4cm}|>{\centering\arraybackslash}m{1cm}|}
\hline
\multicolumn{2}{|c|}{\textbf{Not considering congestion}} & \multicolumn{4}{c|}{\textbf{Considering congestion}} \\
\hline
\textbf{Pricing mechanism} & \textbf{Benefit} & \textbf{Pricing mechanism - Global cases} & \textbf{Benefit} & \textbf{Pricing mechanism - Local cases} & \textbf{Benefit} \\
\hline
SP & 7617 & SP (A\&B) & 7617 & SP (A\&B) & 7617 \\
\hline
TP & 3636 ($\downarrow$52\%) & SP in normal case, \newline TP (A\&B) when congestion & 6380 ($\downarrow$16\%) & SP in normal case, \newline SP (B), TP (A) when congestion & 6998
($\downarrow$8\%)\\
\hline
DP & 4019 ($\downarrow$47\%) & SP in normal case, \newline DP (A\&B) when congestion & 6380 ($\downarrow$16\%) & SP in normal case, \newline SP (B), DP (A) when congestion & 6998 ($\downarrow$8\%) \\
\hline
\end{tabular}
\label{tab3}
\end{table}

\subsection{Quantitative analysis for the whole year}
An analysis over the course of one year was conducted to quantify both the occurrence and intensity of the unexpected peak flow issue, as well as to evaluate the efficiency of the proposed pricing mechanisms. The results, summarized in Table~\ref{tab4}, indicate that the unexpected peak flow issue due to simultaneous reactions from the BRP may arise for approximately 86 hours per year (344 ISPs). Implementing the dual-price mechanism reduces the number of such intervals by nearly 15\%, which is more effective than the two-price mechanism, which achieves a 11.9\% reduction. Additionally, the new pricing mechanism resolves about 19.1\% to 20.2\% of the days in which congestion occurs in case using the two-price and the dual-price mechanisms.

\begin{table}[htbp]
\centering
\caption{Frequency of congestion by different IPMs}
\fontsize{8.5}{12}\selectfont
\begin{tabular}{|m{2cm}|m{2.5cm}|m{2.8cm}|m{2.8cm}|}
\hline
\multirow{2}{*}

\textbf{Indicator} & 
\textbf{Single price } & 
\textbf{Dual price } &
\textbf{Two price }
\\
\hline
        \textbf{ISP} & 344 & 292 ($\downarrow$ 15.1\%) & 303 ($\downarrow$ 11.9\%) \\
\hline
        \textbf{Day} & 94 & 75 ($\downarrow$ 20.2\%) & 76 ($\downarrow$ 19.1\%) \\
\hline
        \textbf{Week} & 35 & 31 ($\downarrow$ 11.4\%) & 33 ($\downarrow$ 5.7\%)\\
\hline
        \textbf{Total hour} & 86 & 73 ($\downarrow$ 15.1\%) & 75.75 ($\downarrow$ (11.9\%)\\
\hline
\end{tabular}
\label{tab4}
\end{table}

\begin{figure}[htbp]
\centerline{\includegraphics[width=0.5\columnwidth]{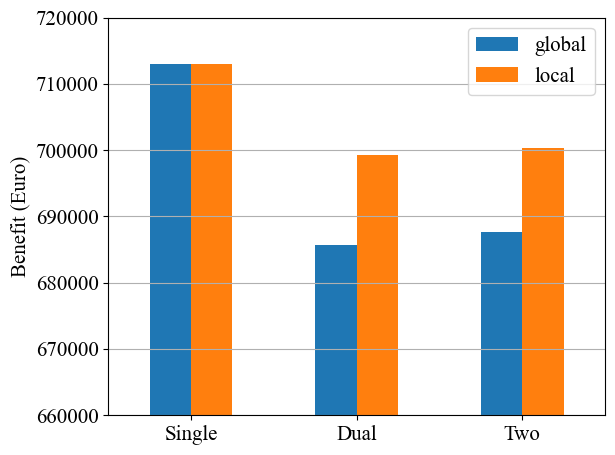}}
\caption{BRP's profit of the whole year in different cases.}
\label{fig: benefit}
\end{figure}

The results show that the number of congestion ISPs is reduced more under the dual-price mechanism than the two-price mechanism.   
It means more passive balancing action is eliminated by the dual-price, leading to a reduction of greater potential benefit compared to the two-price mechanism. 
Due to the number of ISPs being solved by the two-price mechanism being less than dual-price, it means more benefit from passive balancing is still secure, leading to more benefit gained by the two-price mechanism than the dual-price mechanism. The comparison of the benefits in different cases is demonstrated in Fig.~\ref{fig: benefit}.
A reduction of 3.6–3.8\% in the annual benefit is observed when the new IPM is applied globally, whereas only a 1.8–1.9\% reduction is observed when these new IPMs are applied locally.
Consensus with the previous result, the calculation for the whole year proves that applying locally the new price mechanism helps maintain the incentive for the BRP in the non-congested area, which explains the higher benefit than applying globally.
The single-price combined with the two-price mechanism yields a higher overall benefit than combined with the dual-price mechanism because the dual-price mechanism only generates profit from surplus situations, while shortages, multiplied by the zero upward regulation price, do not yield any gain. 
The two-price mechanism offers greater financial benefits but achieves a lower frequency of congestion resolution, whereas the dual-price mechanism is more effective at resolving congestion, in turn reducing the profits. 
The congestion cost or maintenance cost, or the priority of the system operator, are the indicators to select a suitable replacement pricing mechanism.

This study currently focuses on a BRP model based on EV fleets as an example and does not yet account for the potential of vehicle-to-grid (V2G) capabilities or other types of BRPs. As a result, the outcomes may not fully capture the broader range of behaviors and impacts in a realistic setting. Future work will address this limitation by incorporating additional devices and exploring their implementation in real-world applications.
Another reflection is that, due to the difficulty of accurately forecasting imbalance prices and regulation states, BRPs may act more cautiously when providing passive balancing. This increased caution could, in practice, reduce the likelihood of congestion. Moreover, in the Netherlands, the temporary delay in publishing imbalance information has led to a rapid increase in the occurrence of regulation state 2. As a result, BRPs now perceive passive balancing as riskier than before, which underscores the importance of considering the risk-averse behavior of BRPs in future analysis.
Additionally, other research directions include examining the competitive behavior among BRPs, determining optimal budget allocation for a service provider acting as both BRP and BSP, and analyzing their behavior under imperfect information.

\section{Conclusion and policy implications}
\label{conclusion}
In this study, we investigate a peak-flow issue in the Netherlands caused by BRPs’ passive balancing behavior. To address this problem, we propose a congestion-dependent pricing mechanism designed to mitigate the unexpected peak flow while preserving the system operator’s incentive to maintain system balance for the BRPs in non-congested areas. A proof of concept is demonstrated using a two-stage BRP model and three imbalance pricing mechanisms: the single-price, dual-price, and two-price mechanisms, while accounting for uncertainties in day-ahead and imbalance prices.
The results show that the single-price mechanism yields the highest BRP profit by enabling passive balancing contributions. In contrast, the two-price and dual-price mechanisms substantially reduce BRP revenues by limiting passive balancing opportunities. However, the strong incentives of the single-price mechanism can exacerbate the unexpected peak flow problems in the network.
With the proposed mechanism, nearly 15\% of peak flows caused by the overreaction of BRPs with the single-price incentive are eliminated in the entire year. In return, BRP profits are modestly reduced by only 1.8\%, which is considered acceptable when compared to the potential costs of grid congestion and infrastructure degradation. Furthermore, a year-long quantitative analysis highlights the trade-off between congestion mitigation and BRP profitability when using the dual-price or two-price mechanisms.

Nowadays, the unexpected peak flow at the TSO-DSO interface occurs frequently, driven by the TSO activation of balancing service or BRP forecasting errors and passive balancing overreactions. The proposed pricing mechanism offers a promising solution from the BRP perspective by maintaining their profitability while increasing their awareness of the grid conditions, without requiring major changes to the overall system design.
Based on these findings and our earlier work, we formulate two recommendations to policymakers:
\begin{itemize}
\item Enhancing the TSO-DSO coordination by implementing the reserve market merit order check with the DSO, as described in~\cite{26}.
\item Implement the congestion-dependent imbalance pricing mechanism as proposed in this paper.
\end{itemize}

Naturally, implementing these recommendations requires time for transition and needs to consider the potential barriers.
A prerequisite is to gain agreement from the DSOs, BRPs, and TSO on data sharing and acceptance of settlement under the new imbalance pricing mechanism. Congestion state information should be communicated by each DSO to the BRPs connected to its grid, while regional e-programs should be provided to the TSO by BRPs rather than in an aggregated form. Using the congestion state data and the regional e-program, the TSO settles each BRP using the appropriate pricing mechanism. It should be noted that this approach increases the complexity of BRP operation, since splitting their portfolios into smaller regions may lead to larger imbalance errors. In addition, the TSO is required to perform an additional step to determine the appropriate IPM for each ISP. Improvements in forecasting the imbalance prices and system regulation states are essential to support the effective implementation of this mechanism. Therefore, to evaluate the efficiency, not only the reductions in the unexpected peak flows, frequency fluctuations, BRP imbalance, congestion costs and changes in BRP behavior, but also Information Technology (IT) costs for the additional data transfer and infrastructure upgrades should be taken into account.

Reflecting on the current regulation, the primary IPM of the proposed mechanism is the single-price system, which aligns with the current practice and the ACER (2020) harmonization document as well as the EU Balancing Guideline. These frameworks promote harmonization while allowing flexibility for national congestion mitigation. Meanwhile, TSOs are granted a significant degree of freedom to propose their own methodologies. In this context, the proposal to apply a two-price or dual-price mechanism specifically in congested areas remains consistent with the intent of these frameworks.

However, this adjustment may raise fairness concerns, as BRPs in different regions would be treated differently. Furthermore, the conditions under which the DSO can notify a BRP that its region is congested — whether before or after considering other congestion management services — also raise questions of fairness. Some distortion may arise between congested and non-congested areas, which must be evaluated by the engaged stakeholders.
The suitable alternative IMP in congested areas should be carefully evaluated based on the operational and economic efficiency. 
The exploration of the BRP model with different assets and the feasibility of implementing the proposed approach in real-world applications, as well as the risk-averse behavior of BRPs, will be investigated in future work.

\section{Aknowledgement}
The authors gratefully acknowledge the financial support provided by the NWO-funded DEMOSES project. Additionally, the authors would like to thank Nguyen Huu Thien An from the Electrical Energy Systems Group at Eindhoven University of Technology for his insightful discussion throughout the course of this research. 
\appendix
\section{Appendix}
\label{appendix}
\subsection{Virtual battery data}
The upper and lower charging paths are calculated by Eq.~\ref{eq:charge_max} and \ref{eq:charge_min}. $\overline{E^{EV}_t}$ and $\underline{E^{EV}_t}$ present the maximum and minimum values of the energy of the VB, which is the total of the charging path of all EVs and the additional energy from the EVs returning home ($E^{arr}_{t}$), minus the energy reduced from the EVs leaving home ($E^{dep}_t$) at time t. The minimum energy boundary is formed by the ALAP charging path ($E^{ASAP}_{n,t}$), while the ASAP charging path ($E^{ALAP}_{n,t}$) is used to define the maximum energy boundary. $N^{arr}_t$ and $N^{dep}_t$ are the number of EVs arriving and departing at time t, respectively.

\begin{align}
\overline{E^{EV}_t} = \sum_{n}^{N} E^{ASAP}_{n,t} + E^{arr}_t - E^{dep}_t &\quad \forall t \label{eq:charge_max}\\
\underline{E^{EV}_t} = \sum_{n}^{N} E^{ALAP}_{n,t} + E^{arr}_t - E^{dep}_t &\quad \forall t \label{eq:charge_min}\\
E^{arr/dep}_t = \sum_{n}^{N^{arr/dep}_t} E^{arr/dep}_{n,t} &\quad \forall t 
\end{align}

\subsection{Constraints in the DA stage with Single price mechanism}
\begin{align}
E^{EV}_t = E^{init} + E^{arr}_t - \eta E^{da}_{t} &\quad \forall t = 0 \label{eq:EEV0}\\
E^{EV}_{t} = E^{EV}_{t-1} + E^{arr}_t - E^{dep}_{t-1} - \eta E^{da}_t &\quad\forall t > 0 \label{eq:EEVt}\\
\underline{E^{EV}_t} \leq E^{EV}_t \leq 
\overline{E^{EV}_t} &\quad\forall t \label{eq:EDA_plim}\\
p_{charge_t} \leq  E^{da}_t /\Delta t \leq 0 &\quad\forall t \label{eq:EDA_lim}\\
p_{dis_t} = -p_{charge_t} = \sum_{k}^{N_{parked}} P^{\max}_k &\quad \forall t 
\end{align}

The maximum charging power of the VB at time $t$, $p_{charge_t}$, is obtained by summing up all the EVs parked at home ($N_{parked}$). In which $P^{\max}_k$ is the maximum charging rate of EV number $k$.
Based on the energy bought from the market and the total energy increase or decrease due to the EV arriving and departing, the total energy of the virtual battery is determined by Eq.~\ref{eq:EEV0} and Eq.~\ref{eq:EEVt}. DA price is forecasted with a certain level of accuracy through different scenarios. 
$E^{da}_t$ is the amount of energy that will be bought from the DA market at time t in the next day. This value is limited by the maximum charging rate in Eq.~\ref{eq:EDA_plim} and the maximum/minimum boundaries as described in Eq.~\ref{eq:EDA_lim}.
$\eta$ is the charging efficiency of the EVs, $\Delta t$ is the duration of one PTU/ISP in an hour.

\subsection{Constraints in the real-time stage}
\begin{align}
-p^{charge}_t \leq E^{rt}_{t}/ \Delta t \leq 0 &\quad\forall t\\
\Delta E^{\uparrow}_t + \Delta E^{\downarrow}_t = E^{rt}_{t} - E^{da}_t &\quad\forall t \label{eq:EVrt_cal}\\
-p_{charge_t} \leq  (E^{da}_t + \Delta E^{\downarrow}_{t}) /\Delta t \leq 0 &\quad\forall t\\
-p_{charge_t} \leq  E^{rt}_t /\Delta t \leq 0 &\quad\forall t\\
\underline{E^{EV}_t} \leq E^{EV}_t \leq 
\overline{E^{EV}_t} &\quad\forall t\\
-p_{charge_t}u_t \leq  \Delta E^{\downarrow}_{t} /\Delta t \leq 0 &\quad\forall t \label{eq:Edn_rt}\\
0 \leq  \Delta E^{\uparrow}_{t} /\Delta t \leq p_{charge_t}(1-u_t) &\quad\forall t \label{eq:Eup_rt}\\
E^{EV}_t = E^{init} + E^{arr}_t - \eta (E^{da}_{t} + \Delta E^{\downarrow}_{t})
- \Delta E^{\uparrow}_{t} \quad \forall t= 0 \label{eq:EEV0_rt}\\
E^{EV}_{t} = E^{EV}_{t-1} + E^{arr}_t - E^{dep}_{t-1} - \eta (E^{da}_{t} + \Delta E^{\downarrow}_{t})  
- \Delta E^{\uparrow}_{t} \quad\forall t > 0 \label{eq:EEVt_rt}
\end{align}

The energy in real-time is determined by the $\Delta E^{\uparrow}_t$, $\Delta E^{\downarrow}_t$ and the $E^{da}_t$ in Eq.~\ref{eq:EVrt_cal}. Similar to Eq.~\ref{eq:EEV0} and \ref{eq:EEVt}, Eq.~\ref{eq:EEV0_rt} and \ref{eq:EEVt_rt} show the total energy of VB at time step t and its relationship with the state in the previous time step. A binary variable $u$ is used in Eq.~\ref{eq:Edn_rt} and Eq.~\ref{eq:Eup_rt} to help prevent the surplus and shortage from happening at the same time.

\bibliography{references}
\end{document}